\documentclass[pra,aps,twocolumn,aps,superscriptaddress]{revtex4-1}
\usepackage{times}
\usepackage{graphicx,afterpage}
\usepackage{epsfig}
\usepackage{amsfonts}
\usepackage{amsmath}
\usepackage{amssymb}
\usepackage{color}
\usepackage{multirow}
\usepackage[colorlinks=true,citecolor=blue,urlcolor=blue]{hyperref}
\usepackage[T2A,T1]{fontenc}
\usepackage[utf8]{inputenc}
\usepackage[russian,english]{babel}
\usepackage{booktabs}
\usepackage{array} 
\newcolumntype{C}{>{$}c<{$}} 
\newcolumntype{L}{>{$}l<{$}} 

\begin{document}


\title{Maximum nonlocality in the (3,2,2) scenario}




\author{Sheila L\'opez-Rosa}
\email{slopezrosa@us.es}
\affiliation{Departamento de F\'{\i}sica Aplicada II,
Universidad de Sevilla,
E-41012 Sevilla, Spain}

\author{Zhen-Peng Xu}
\email{zhenpengxu@us.es}
\affiliation{Theoretical Physics Division, Chern Institute of Mathematics,
Nankai University,
Tianjin 300071, People's Republic of China}
\affiliation{Departamento de F\'{\i}sica Aplicada II,
Universidad de Sevilla,
E-41012 Sevilla, Spain}

\author{Ad\'an~Cabello}
\email{adan@us.es}
\affiliation{Departamento de F\'{\i}sica Aplicada II,
Universidad de Sevilla,
E-41012 Sevilla, Spain}


\begin{abstract}
We identify the simplest combinations of entanglement and incompatibility giving the maximum quantum violation for each of the 46 classes of tight Bell inequalities for the (3,2,2) scenario, i.e., three parties, two measurements per party, and two outcomes per measurement. This allows us to classify the maximum quantum nonlocality according to the simplest resources needed to achieve it. We show that entanglement and incompatibility only produce maximum nonlocality when they are combined in specific ways. For each entanglement class there is, in most cases, just one incompatibility class leading to maximum nonlocality. We also identify two interesting cases. We show that the maximum quantum violation of \'Sliwa inequality 23 only occurs when the third party measures the identity, so nonlocality cannot increase when we add a third party to the bipartite case. Almost quantum correlations predict that adding a new party increases nonlocality. This points out that either almost quantum correlations violate a fundamental principle or that there is a form of tripartite entanglement which quantum theory cannot account for. The other interesting case is the maximum quantum violation of \'Sliwa inequality 26, which, like the Mermin inequality, requires maximum incompatibility for all parties. In contrast, it requires a specific entangled state which has the same tripartite negativity as the W state.
\end{abstract}



\maketitle


\section{Motivations}


\subsection{How entanglement and incompatibility combine for producing maximum nonlocality}


Entanglement and incompatibility are the two basic ingredients for nonlocality (i.e., violation of Bell inequalities \cite{Bell64}). Commonly, their significance for nonlocality has been studied separately. In the literature, there are works on the relation between entanglement and nonlocality \cite{Gisin91,PR92,Werner89,AGT06,Vertesi08,BFFHB16,CGRS16,HQVPB16}, and other works on the relation between incompatibility and nonlocality \cite{KC85,WPF09,QBHB16}.
However, arguably, nonlocality is a consequence of how entanglement and incompatibility work {\em together}. Little is known about how they should be {\em combined} to produce maximum nonlocality, defined as maximum violation of tight Bell inequalities. Most of what we know can be summarized as follows.


\subsubsection{$(2,2,2)$ scenario}


The maximum quantum violation of the only nontrivial tight Bell inequality in the two-party, two-setting, two-outcome or $(2,2,2)$ scenario, the Clauser-Horne-Shimony-Holt (CHSH) inequality \cite{CHSH69}, is only achieved when both maximum entanglement and maximum incompatibility between the local observables of both parties concur \cite{SW87,Tsirelson93}. We say that two quantum projective $d$-outcome local measurements $A$ and $a$ are maximally incompatible (or maximally value complementary \cite{BL95}) when $|\langle a=i | A=j \rangle |^2 = \frac{1}{d}$, for all $i,j$, where $|A=j\rangle$ denotes the eigenstate of $A$ with outcome $j$. Nonmaximal incompatibility can be quantified in several ways \cite{HMZ16}. We will later introduce a particularly suitable quantifier for it.


\subsubsection{$(2,3,2)$ scenario}


For the $(2,3,2)$ scenario, i.e., the two-party, three-setting, two-outcome scenario, the maximum quantum violation of the only nontrivial tight Bell inequality, the $I_{3322}$ inequality \cite{Froissart81,CG04}, cannot be achieved with qubits and it is conjectured to require local systems of infinite dimension \cite{PV10}. However, it is known that the largest violation is not obtained by the maximally entangled state, even if its dimension is allowed to be arbitrarily large \cite{VW11}. The maximum violation with qubits is achieved with a maximally entangled state and nonmaximally incompatible local measurements with the same structure of incompatibility for both parties. Specifically, $|\langle a=i | A=j \rangle |^2$ is either $1/4$ or $3/4$ \cite{CG04}.


\subsubsection{$(2,2,d)$ scenario}


For the $(2,2,d)$ scenario, with $d>2$, the maximum quantum violation of the most famous family of tight Bell inequalities, the Collins-Gisin-Linden-Massar-Popescu inequalities \cite{CGLMP02}, cannot be attained with maximum incompatibility or maximum entanglement \cite{ADGL02,ZG08}. For example, for $d=3$, the maximum violation occurs for the state
$|\psi_\rangle = \frac{1}{n} \left(|00\rangle + \gamma|11\rangle+|22\rangle\right)$,
where $\gamma = (\sqrt{11}-\sqrt{3})/2$ and $n=2+\gamma^2$ and the local measurements are not maximally incompatible although they have the same structure of incompatibility for both parties. Specifically, $|\langle a=i | A=j \rangle |^2$ is either $1/9$ or $4/9$ \cite{ADGL02}.
For increasing $d$, the entanglement entropy of the state leading to the maximum nonlocality increases but not as fast as for the maximally entangled state, indicating that the state leading to the maximum nonlocality is progressively separating from the maximally entangled state \cite{ZG08}. Simultaneously, in the optimal local measurements, the number of different terms in the set $|\langle A_2 =j | A_1=k \rangle |^2$ grows with $d$. One can construct a nontight Bell inequality which is maximally violated by the same local observables, but using maximally entangled states \cite{SLK04}.


\subsubsection{Other scenarios}


Greenberger-Horne-Zeilinger (GHZ) states \cite{GHZ89} and $n$-qubit graph states, together with maximally incompatible measurements, maximally violate Bell inequalities (some of them tight) in the $(n,2,2)$ scenario \cite{Mermin90,Ardehali92,BK93} and in scenarios with $n$ parties in which each party has two or three measurements, all of them with two outcomes \cite{GC08,CGR08}. For most of these scenarios the number of inequivalent classes of tight Bell inequalities is unknown.


\subsubsection{$(3,2,2)$ scenario}


From the previous examples, it is hard to extract any general conclusion about which combinations of entanglement and incompatibility lead to maximum quantum nonlocality for tight Bell inequalities. However, there is a scenario which may help for this purpose and which is also interesting for other reasons. An analysis of which combinations of entanglement and incompatibility are needed to reach the quantum maxima is presented in this paper for this scenario. We are talking about the $(3,2,2)$ scenario.

The first reason why this scenario is interesting is because there are exactly 46 inequivalent classes of tight Bell inequalities rather than just one or an unknown number of them, as in the case of other scenarios. These 46 classes were first obtained by Pitowsky and Svozil \cite{PS01} and \'Sliwa \cite{Sliwa03}. For convenience, we list in Table \ref{Table1} the tight Bell inequalities representing each of these classes, in the version of \'Sliwa \cite{Sliwa03}. This is useful, specially since none of the published versions of \cite{PS01,Sliwa03} contains them.
This variety of cases provides the opportunity to investigate which specific combinations of entanglement and incompatibility produce maximum nonlocality and, in this way, enables us to start understanding how they should complement each other. In addition, this analysis also allows us to classify tripartite quantum maximum nonlocality according to the two resources that produce it.


\begin{widetext}
\begin{turnpage}
\begin{table*}
\vspace*{-10mm}
\begin{center}
\caption{\small Tight Bell inequalities for the $(3,2,2)$ scenario or \'Sliwa inequalities \cite{Sliwa03}. $A$ and $a$ are the measurements of the first party, $B$ and $b$ of the second, and $C$ and $c$ of the third. Measurements outcomes are $1$ and $-1$. $Abc$ denotes the average $\langle Abc \rangle$. The numbering of the classes and the choice of representant for each class is the same as in \cite{Sliwa03}.}
\label{Table1}
\scalebox{1.00}{
\hspace*{-18mm}
\begin{tabular}{ccl}
\hline \hline
\'Sliwa class & Local maximum & Bell operator \\
\hline
$1$  & $1$ & $ A  +  B  -  AB  +  C  -  AC  -  BC  +  ABC $\\
$2$  & $2$ & $ ABC  +  abC  +  aBc  -  Abc $\\
$3$  & $2$ & $ ABC  +  aBC  +  Abc  -  abc $\\
$4$  & $2$ & $2 A  +  BC  -  ABC  +  bC  -  AbC  +  Bc  -  ABc  -  bc  +  Abc $\\
$5$  & $3$ & $ A  +  B  +  aB  +  Ab  -  ab  +  C  +  aC  -  ABC  -  aBC  +  bC  -  AbC  +  Ac  -  ac  +  Bc  -  ABc  -  bc  +  abc $\\
$6$  & $3$ & $ A  +  B  +  AB  +  C  +  aC  -  ABC  -  aBC  +  bC  -  AbC  +  Ac  -  ac  -  Bc  +  aBc  +  bc  -  Abc $\\
$7$  & $4$ & $3 ABC  +  aBC  +  AbC  -  abC  +  ABc  -  aBc  -  Abc  +  abc $\\
$8$  & $4$ & $ AB  +  aB  +  Ab  +  ab  + 2 ABC  - 2 abC  +  ABc  -  aBc  -  Abc  +  abc $\\
$9$  & $4$ & $ AB  +  aB  +  Ab  +  ab  + 2 ABC  - 2 AbC  +  ABc  -  aBc  +  Abc  -  abc $\\
$10$ & $4$ & $ AB  +  aB  +  Ab  +  ab  +  AC  -  aC  +  BC  +  ABC  -  bC  -  abC  +  Ac  -  ac  -  Bc  +  aBc  +  bc  -  Abc $\\
$11$ & $4$ & $2 AB  + 2 ab  +  ABC  +  aBC  -  AbC  -  abC  +  ABc  -  aBc  +  Abc  -  abc $\\
$12$ & $4$ & $2 AB  + 2 ab  +  AC  +  aC  -  BC  +  aBC  -  bC  -  AbC  +  Ac  +  ac  -  Bc  -  aBc  -  bc  +  Abc $\\
$13$ & $4$ & $2 AB  + 2 aB  +  ABC  -  aBC  +  AbC  -  abC  +  ABc  -  aBc  -  Abc  +  abc $\\
$14$ & $4$ & $2 AB  + 2 aB  +  AC  -  aC  +  AbC  -  abC  +  Ac  -  ac  -  Abc  +  abc $\\
$15$ & $4$ & $2 AB  + 2 aB  +  AC  +  aC  - 2 BC  +  AbC  -  abC  +  Ac  +  ac  - 2 Bc  -  Abc  +  abc $\\
$16$ & $4$ & $ A  +  a  +  AB  +  aB  +  AC  +  aC  - 2 aBC  +  AbC  -  abC  +  ABc  -  aBc  -  Abc  +  abc $\\
$17$ & $4$ & $ A  +  a  +  AB  +  aB  +  AC  +  aC  -  ABC  -  aBC  + 2 Abc  - 2 abc $\\
$18$ & $4$ & $ A  +  a  +  AB  +  aB  +  AC  +  aC  - 2 BC  +  AbC  -  abC  +  ABc  -  aBc  + 2 bc  -  Abc  -  abc $\\
$19$ & $4$ & $ A  +  a  +  AB  +  aB  +  AC  +  aC  - 2 BC  + 2 bC  -  AbC  -  abC  +  ABc  -  aBc  +  Abc  -  abc $\\
$20$ & $4$ & $ A  +  a  +  AB  -  aB  +  Ab  -  ab  +  AC  -  aC  -  BC  +  ABC  +  aBC  -  bC  +  AbC  +  abC  +  Bc  -  ABc  -  aBc  -  bc  +  Abc  +  abc $\\
$21$ & $4$ & $ A  +  a  +  B  +  AB  +  b  -  ab  +  AC  +  aC  +  BC  - 2 ABC  -  aBC  +  bC  -  AbC  +  ABc  -  aBc  -  Abc  +  abc $\\
$22$ & $4$ & $ A  +  a  +  B  +  AB  +  b  -  ab  +  C  +  AC  +  BC  - 2 ABC  -  aBC  -  AbC  +  abC  +  c  -  ac  -  ABc  +  aBc  -  bc  +  Abc $\\
$23$ & $4$ & $ A  +  a  +  B  -  AB  -  aB  +  b  -  Ab  -  ab  +  AC  -  aC  -  ABC  +  aBC  -  AbC  +  abC  +  Bc  -  ABc  -  aBc  -  bc  +  Abc  +  abc $\\
$24$ & $5$ & $ A  +  B  +  aB  +  Ab  +  ab  +  C  +  aC  -  BC  + 2 ABC  -  aBC  - 2 abC  +  Ac  +  ac  - 2 aBc  -  Abc  +  abc $\\
$25$ & $5$ & $ A  +  B  +  aB  +  Ab  +  ab  +  C  +  aC  -  BC  + 2 ABC  -  aBC  - 2 abC  +  Ac  +  ac  - 2 ABc  +  Abc  -  abc $\\
$26$ & $5$ & $ A  +  B  +  AB  + 2 ab  +  C  +  AC  +  BC  -  ABC  - 2 abC  + 2 ac  - 2 aBc  - 2 bc  + 2 Abc $\\
$27$ & $5$ & $2 A  +  a  +  B  -  AB  +  Ab  +  ab  +  C  -  AC  + 2 ABC  - 2 aBC  +  bC  -  AbC  +  Ac  +  ac  +  Bc  -  ABc  +  bc  - 2 Abc  -  abc $\\
$28$ & $6$ & $ A  +  a  +  AB  -  aB  +  AC  -  aC  -  BC  + 2 ABC  +  aBC  +  bC  -  AbC  - 2 abC  +  Bc  -  ABc  - 2 aBc  +  bc  - 3 Abc $\\
$29$ & $6$ & $ A  +  a  +  AB  -  aB  +  AC  -  aC  -  BC  + 2 ABC  +  aBC  +  bC  -  AbC  - 2 abC  +  Bc  - 3 ABc  +  bc  -  Abc  - 2 abc $\\
$30$ & $6$ & $ A  +  a  + 2 AB  - 2 aB  +  Ab  -  ab  +  AC  -  aC  -  BC  + 2 ABC  +  aBC  -  bC  +  AbC  + 2 abC  +  Bc  - 2 ABc  -  aBc  -  bc  + 2 Abc  +  abc $\\
$31$ & $6$ & $ A  +  a  +  B  -  aB  +  b  -  Ab  +  AC  -  aC  + 2 aBC  -  AbC  + 3 abC  +  Bc  - 2 ABc  -  aBc  -  bc  + 2 Abc  +  abc $\\
$32$ & $6$ & $ A  +  a  +  B  -  aB  +  b  -  Ab  + 2 AC  - 2 aC  + 2 aBC  + 2 abC  +  Ac  -  ac  -  Bc  + 2 ABc  +  aBc  +  bc  -  Abc  - 2 abc $\\
$33$ & $6$ & $ A  +  a  +  B  -  aB  +  b  -  Ab  +  C  -  aC  + 2 aBC  -  bC  + 2 AbC  +  abC  +  c  -  Ac  -  Bc  + 2 ABc  +  aBc  +  Abc  - 3 abc $\\
$34$ & $6$ & $ A  +  a  +  B  -  aB  +  b  -  Ab  +  C  -  aC  -  BC  - 2 ABC  +  aBC  - 2 bC  + 2 AbC  + 2 abC  +  c  -  Ac  - 2 Bc  -  bc  +  Abc  - 2 abc $\\
$35$ & $6$ & $ A  +  a  +  B  -  AB  - 2 aB  +  b  - 2 Ab  -  ab  +  AC  -  aC  -  ABC  +  aBC  - 2 AbC  + 2 abC  +  Bc  - 2 ABc  -  aBc  -  bc  + 2 Abc  +  abc $\\
$36$ & $6$ & $2 A  +  AB  +  aB  +  Ab  +  ab  +  AC  +  aC  +  BC  - 2 ABC  +  aBC  -  bC  +  AbC  - 2 abC  +  Ac  +  ac  -  Bc  +  ABc  - 2 aBc  -  bc  + 2 Abc  -  abc $\\
$37$ & $6$ & $2 A  +  AB  +  aB  +  Ab  +  ab  +  AC  +  aC  +  BC  - 3 ABC  -  bC  + 2 AbC  -  abC  +  Ac  +  ac  -  Bc  + 2 ABc  -  aBc  -  bc  +  Abc  - 2 abc $\\
$38$ & $6$ & $2 A  + 2 AB  + 2 aB  +  AC  +  aC  -  BC  +  ABC  - 2 aBC  +  bC  - 2 AbC  +  abC  +  Ac  +  ac  -  Bc  +  ABc  - 2 aBc  -  bc  + 2 Abc  -  abc $\\
$39$ & $6$ & $2 A  + 2 B  -  AB  +  aB  +  Ab  +  ab  + 2 C  -  AC  +  aC  -  BC  + 2 ABC  -  aBC  +  bC  -  AbC  - 2 abC  +  Ac  +  ac  +  Bc  -  ABc  - 2 aBc  +  bc  - 2 Abc  +  abc $\\
$40$ & $6$ & $2 A  + 2 a  + 2 B  -  AB  -  aB  +  Ab  +  ab  +  AC  +  aC  + 2 BC  -  ABC  -  aBC  + 2 bC  - 2 AbC  - 2 abC  +  Ac  -  ac  - 2 ABc  + 2 aBc  +  Abc  -  abc $\\
$41$ & $7$ & $ A  +  B  +  AB  +  C  +  aC  - 3 ABC  -  aBC  +  bC  -  AbC  - 2 abC  +  Ac  -  ac  +  Bc  - 4 ABc  +  aBc  -  bc  +  Abc  + 2 abc $\\
$42$ & $8$ & $ A  +  a  +  B  +  AB  +  b  -  ab  +  AC  -  aC  +  BC  - 2 ABC  -  aBC  -  bC  -  AbC  + 4 abC  + 2 ac  -  ABc  - 3 aBc  + 2 bc  - 3 Abc  -  abc $\\
$43$ & $8$ & $2 A  + 2 B  -  AB  +  aB  +  Ab  -  ab  +  AC  +  aC  +  BC  - 2 ABC  - 3 aBC  -  bC  +  AbC  + 2 abC  +  Ac  -  ac  +  Bc  - 3 ABc  +  bc  - 4 Abc  +  abc $\\
$44$ & $8$ & $2 A  + 2 a  + 2 AB  - 2 aB  +  AC  -  aC  - 2 BC  + 2 ABC  + 2 aBC  + 2 bC  -  AbC  - 3 abC  +  Ac  -  ac  - 2 Bc  + 2 ABc  + 2 aBc  - 2 bc  + 3 Abc  +  abc $\\
$45$ & $8$ & $3 A  +  a  + 2 AB  - 2 aB  +  Ab  -  ab  + 2 AC  - 2 aC  - 2 BC  + 2 ABC  + 2 aBC  - 2 bC  + 2 AbC  + 2 abC  +  Ac  -  ac  - 2 Bc  + 2 ABc  + 2 aBc  + 2 bc  - 3 Abc  -  abc $\\
$46$ & $10$ & $3 A  +  a  + 3 B  - 2 AB  -  aB  +  b  -  Ab  - 2 ab  + 2 AC  - 2 aC  +  BC  - 3 ABC  + 4 aBC  +  bC  -  AbC  + 2 abC  +  Ac  +  ac  + 2 Bc  - 3 ABc  -  aBc  - 2 bc  + 4 Abc  + 2 abc $\\
\hline \hline
\end{tabular}}
\end{center}
\end{table*}
\end{turnpage}
\end{widetext}


Second, for the $(3,2,2)$ scenario, it has been proven that all the quantum maxima (i.e., all the maximum violations of the tight Bell inequalities) are attained by measuring projective observables on three-qubit pure states \cite{Masanes03}. This implies that there is a good chance to obtain analytical expressions for the quantum maxima and the corresponding states and measurements (rather than just numerical ones).

A third reason, related to the previous one, is that three-qubit entanglement is much richer than two-qubit entanglement, but still simple enough so that we can provide an exhaustive and manageable classification of the types of entanglement, while this is not the case for the entanglement needed for more complex Bell inequality scenarios.


\subsection{The principle of quantum correlations}


However, perhaps the main reason that justifies the interest of our analysis is the fact that most proposed principles for quantum correlations have been proven to fail to explain correlations in the $(3,2,2)$ scenario. Information causality \cite{PPKSWZ09} fails to explain the impossibility of specific $(3,2,2)$ nonlocal nonsignaling correlations \cite{GWAN11,YCATS12}. Triviality of communication complexity \cite{vanDam99,vanDam13}, macroscopic locality \cite{NW09}, and local orthogonality \cite{FSABCLA13} also fail, as there are $(3,2,2)$ nonquantum correlations that satisfy all these principles (as shown here and also in \cite{VSL16}). These correlations are called almost quantum \cite{NGHA15}. Consequently, any principle which explains the quantum maxima in the $(3,2,2)$ scenario has good chances to be {\em the} long awaited principle for quantum correlations. However, how can we expect to identify that principle if we do not know how quantum theory manages to achieve these maxima? The purpose of this work is precisely to address this problem.


\section{Methods}


\subsection{Maximum nonlocality: States and measurements}


For calculating the quantum maxima for the 46 tight Bell inequalities we use two methods. In the first place, we take advantage of the fact that the quantum maxima can be achieved with pure states of three qubits and projective measurements \cite{Masanes03}. This helps us to try to derive fully analytical results using several standard mathematical programs. We succeeded in 27 out of the 46 cases. We provide an analytical result whenever the analytical maximum matches with, at least, 9 digits the largest of the maxima obtained with numerical optimization methods. In addition, we use the Navascu\'es-Pironio-Ac\'{\i}n (NPA) method \cite{NPA07,NPA08}, up to level $Q_2$, to put numerical upper bounds to the quantum maxima and to compute the maximum for almost quantum correlations. The NPA method has been independently applied to the $(3,2,2)$ scenario by Vallins, Sainz, and Liang \cite{VSL16}.


\subsection{Classification of three-qubit states}


For classifying the entanglement we use a refined version of the classification of three-qubit states proposed by Sab\'{\i}n and Garc\'{\i}a-Alcaine \cite{SG08}. The virtue of the classification in \cite{SG08} with respect to other classifications \cite{AACJLT00,AAJT01} is that it is based on entanglement monotones \cite{Vidal00} which have, by themselves, a direct interpretation as measures of genuine three-qubit entanglement (one of the monotones) and two-qubit entanglement (the other three monotones). The refinement in the classification we introduce here allows us to fully exploit the information about the entanglement given by these monotones and also allows us to single out some entangled states which already had special names before any complete classification was available.

The entanglement monotones needed for our classification are the three bipartite concurrences $C_{AB},C_{AC},C_{BC}$ \cite{HW97} corresponding to the three two-qubit reduced states, and the tripartite negativity $N_{ABC}(\rho)$ of the three-qubit state $\rho$ defined \cite{SG08} as
\begin{equation}
N_{ABC}(\rho)=(N_{A-BC}N_{B-AC}N_{C-AB})^{{1\over3}},
\end{equation}
where the bipartite negativities are defined as $N_{I-JK}=-2\sum_{i}\sigma_{i}(\rho^{TI})$, being $\sigma_{i}(\rho^{TI})$
the negative eigenvalues of $\rho^{TI}$, which is the partial transpose of $\rho$ with respect to subsystem $I$,
$\langle{i_{I},j_{JK}}|{\rho^{TI}}|{k_{I},l_{JK}}\rangle=\langle{k_{I},j_{JK}}|{\rho}|{i_{I},l_{JK}}\rangle$, with $I=A, B, C$,
and $JK=BC, AC, AB$, respectively.
The reasons why $N_{ABC}$ is a better quantifier of fully three-qubit entanglement than other frequently used ones are that $N_{ABC}$ is zero both for fully separable and biseparable states, nonzero for any fully tripartite entangled state, invariant under local unitary transformations, and nonincreasing under local operations and classical communication. Using the values of $N_{ABC},C_{AB},C_{AC}$, and $C_{BC}$, we distinguish among the 12 classes of entanglement shown in Table \ref{Table2}.


\begin{table}[tb]
\begin{center}
\caption{Classification of three-qubit states according to their tripartite and bipartite entanglement. $N_{ABC}$ is the tripartite negativity and $C_{ij},C_{ik},C_{jk}$, with $ \{ i,j,k \} \in \{A,B,C \} $, are the qubit-qubit concurrences, $0<p_n<1$ and $0<q_m<1$. }
\label{Table2}
\begin{tabular}{clcccc}
\hline \hline
Class & Entanglement & $N_{ABC}$ & $C_{ij}$ & $C_{ik}$ & $C_{jk}$ \\
\hline
$0$ & None & $0$ & $0$ & $0$ & $0$ \\
$1$ & 2-qubit nonmax. (Hardy states \cite{Hardy93})& $0$ & $q_0$ & $0$ & $0$ \\
$2$ & 2-qubit maximum (Bell states \cite{BMR92}) & $0$ & $1$ & $0$ & $0$ \\
$3$ & 3-qubit W-like-3 & $p_0$ & $q_1$ & $q_2$ & $q_3$ \\
$4$ & 3-qubit W-like-2 & $p_1$ & $q_4$ & $q_4$ & $q_5$ \\
$5$ & 3-qubit W-like-1 & $p_2$ & $q_6$ & $q_6$ & $q_6$ \\
$6$ & 3-qubit W (W states \cite{DVC00}) & $\frac{2 \sqrt{2}}{3}$ & $\frac{2}{3}$ & $\frac{2}{3}$ & $\frac{2}{3}$ \\
$7$ & 3-qubit star shaped-2 \cite{PB03} & $p_3$ & $q_7$ & $q_8$ & $0$ \\
$8$ & 3-qubit star shaped-1 & $p_4$ & $q_9$ & $q_9$ & $0$ \\
$9$ & 3-qubit 2-1 subtype-1 \cite{SG08} & $p_5$ & $q_{10}$ & $0$ & $0$ \\
$10$ & 3-qubit GHZ-like & $p_6$ & $0$ & $0$ & $0$ \\
$11$ & 3-qubit GHZ (GHZ states \cite{GHZ89}) & $1$ & $0$ & $0$ & $0$ \\
\hline \hline
\end{tabular}
\end{center}
\end{table}


Any three-qubit pure state can always be written as a linear combination of five orthogonal product states \cite{AACJLT00,AAJT01}. On the other hand, the smallest number of orthogonal product states needed for each class in Table \ref{Table2} is known. In fact, for 9 out of the 12 classes, the state can be written with less than five orthogonal product states. Class 0 states just require one product state, classes 1, 2, 10, and 11 just require two orthogonal product states, classes 6 and 9 just require three orthogonal product states, and classes 7 and 8 just require four orthogonal product states. Therefore, identifying the class a state belongs to allows us to write it economically.


\subsection{Classification of incompatibility in the (3,2,2) scenario}


For classifying the incompatibility of each party's measurements $M_1$ and $M_2$, we use a normalized version of the incompatibility monotone $I_{b=0}^{\rm noise} (M_1,M_2)$ introduced by Heinosaari, Kiukas, and Reitzner \cite{HKR15}. Specifically, for quantifying the incompatibility we use $I (M_1,M_2) \equiv (2 + \sqrt{2}) I_{b=0}^{\rm noise} (M_1,M_2)$, which is bounded between $0$, for compatible measurements, and $1$, for maximally incompatible measurements. An incompatibility monotone \cite{HKR15} is zero if and only if the two measurements are compatible, and is nonincreasing under quantum operations. The virtue of this monotone is that it has a direct operational meaning since $I_{b=0}^{\rm noise} (M_1,M_2)$ is the amount of $0$-biased local noise needed to destroy all nonlocal CHSH correlations when added to one observer's measurements (see \cite{HKR15} for details). For projective qubit measurements of the form $M_1=\sigma_x$ and $M_2= \cos \vartheta \sigma_x + \sin \vartheta \sigma_z$,
\begin{equation}
I (M_1,M_2) = (2 + \sqrt{2}) \left[1 - \left(1+\sin \vartheta\right)^{-1/2}\right].
\end{equation}
Based on the values of this incompatibility monotone for the three parties, we distinguish among the 12 classes of incompatibility shown in Table \ref{Table3}.


\begin{table}[tb]
\begin{center}
\caption{Classification of incompatibility in the (3,2,2) scenario based on the incompatibility monotones $I(i_1,i_2)$, $I(j_1,j_2)$, and $I(k_1,k_2)$, with $\{i,j,k \} \in \{A,B,C \}$. $I(i_1,i_2)$ is defined in the text. It is 0 for compatible measurements and 1 for maximally incompatible measurements. ``m'' indicates maximum and ``nm'' indicates nonmaximum, $0<s_n<1$.}
\label{Table3}
\begin{tabular}{clccc}
\hline \hline
Class & Incompatibility & $I(i_1,i_2)$ & $I(j_1,j_2)$ & $I(k_1,k_2)$ \\
\hline
 $0$ & None & $0$ & $0$ & $0$ \\
 $1$ & 2-party nm-2 & $s_0$ & $s_1$ & $0$ \\
 $2$ & 2-party nm-1 & $s_2$ & $s_2$ & $0$ \\
 $3$ & 2-party m, nm & $1$ & $s_3$ & $0$ \\
 $4$ & 2-party m, m & $1$ & $1$ & $0$ \\
 $5$ & 3-party nm-3 & $s_4$ & $s_5$ & $s_6$ \\
 $6$ & 3-party nm-2 & $s_7$ & $s_7$ & $s_8$ \\
 $7$ & 3-party nm-1 & $s_9$ & $s_9$ & $s_9$ \\
 $8$ & 3-party m, nm-2 & $1$ & $s_{10}$ & $s_{11}$ \\
 $9$ & 3-party m, nm-1 & $1$ & $s_{12}$ & $s_{12}$ \\
$10$ & 3-party m, m, nm & $1$ & $1$ & $s_{13}$ \\
$11$ & 3-party m, m, m & $1$ & $1$ & $1$ \\
\hline \hline
\end{tabular}
\end{center}
\end{table}


\section{Maximum nonlocality: Resources needed and classification}


\subsection{Resources needed}


Table \ref{Table4} shows the simplest states and measurements needed to achieve maximum nonlocality for each of the 46 tight Bell inequalities of the $(3,2,2)$ scenario.
In some cases, e.g., in inequalities 2 (which is the Mermin inequality \cite{Mermin90}), 23, and 26, no other combination of resources leads to the maximum nonlocality. In other cases, e.g., in inequalities 3, 6, 13, 14, and 17 there are more options (e.g., using tripartite rather than bipartite entanglement).
We present the simplest of these options, i.e., the one requiring minimal entanglement and incompatibility, as ordered in, respectively, Tables \ref{Table2} and \ref{Table3}.
The following notation is used in Table \ref{Table4}:
\begin{subequations}
\begin{align}
& R(\theta) = \left( \begin{array}{cc}
\cos \theta & -\sin \theta \\
\sin \theta & \cos \theta \end{array} \right), \\
& f=\frac{1}{4} \left(5-\sqrt{17}\right), \\
& s=-\csc ^{-1}\left(\frac{2}{\sqrt{2+\sqrt{78 \sqrt{17}-318}}}\right), \\
& g=160-39 \sqrt{17}, \\
& |\tilde{\pm} \rangle = \frac{1}{\sqrt{2}}\left(|0\rangle \pm i |1 \rangle\right),\\
& |\bar{\pm} \rangle = \frac{1}{\sqrt{2}} \left(-|0\rangle \pm |1\rangle\right), \\
& |\breve{\pm} \rangle = b_{\pm} |0\rangle + c_{\pm} |1 \rangle, \\
& |\hat{\pm} \rangle = \pm d_{\mp} |0 \rangle + d_{\pm} |1 \rangle,
\end{align}
\end{subequations}
where $b_{+} = c_{-} = \frac{1}{\sqrt{2}}$, $b_{-}, c_{+} = \frac{\pm \sqrt{2} - 4 i}{6}$, and $d_{\pm} = \frac{1}{2} \left[2 \pm \sqrt{\frac{1}{2} \left(5 \sqrt{17}-13\right)}\right]^{1/2}$.
Table \ref{Table5} contains the coefficients of some of the states in Table \ref{Table4}.
The quantifiers of entanglement and incompatibility for the states and measurements in Table \ref{Table4} are presented in Table \ref{Table6}.


\begin{widetext}
\begin{turnpage}
\begin{table*}
\vspace*{-8mm}
\begin{center}
\caption{\small Simplest states and measurements needed for maximum nonlocality for the $(3,2,2)$ scenario. ``Maximum'' indicates the maximum quantum violation, $A$ and $a$ are the measurements of the first party, $B$ and $b$ of the second, and $C$ and $c$ of the third. ``None'' indicates that there is no need to prepare any quantum state, $\openone$ indicates that there is no need to perform any measurement, it is enough to always output $1$, $x,z$ denote the corresponding Pauli matrices, $R(\theta), f, s, g,| \tilde{\pm} \rangle, | \bar{\pm} \rangle, |\breve{\pm} \rangle,|\hat{\pm} \rangle$ are defined in Eqs.\ (3), and $\alpha_i,\ldots,\vartheta_i$ are in Table \ref{Table5}.}
\label{Table4}
\scalebox{0.82}{
\hspace*{-30mm}
\begin{tabular}{CCCCCCCCC}
\hline \hline
\text{\'Sliwa} & \text{Maximum} & \text{State} & A & a & B & b & C & c \\
\hline
1 & 1 & \text{None} & \openone & \openone & \openone & \openone & \openone & \openone \\
2 & 4 & \frac{1}{\sqrt{2}} \left( | \tilde{+} \tilde{-} \tilde{-} \rangle + | \tilde{-} \tilde{+} \tilde{+} \rangle \right) & z & x & z & x & z & x\\
3 & 2 \sqrt{2}& \left[R(-3\pi/8) \otimes \openone\right] \frac{1}{\sqrt{2}} \left(|11\rangle_{AB} - |00 \rangle_{AB}\right) & z & x & z & x & -\openone & -\openone\\
4 & 4 \sqrt{2}-2& \left[ R(\pi/8) \otimes \openone \right] \frac{1}{\sqrt{2}} \left(|11\rangle_{BC} - |00 \rangle_{BC}\right) & -\openone & -\openone & z & x & z & x\\
5 & 8 \sqrt{5}-13 & \alpha_5 |\bar{+}\bar{+}\bar{+}\rangle + \beta_5 |\bar{+}\bar{+}\bar{-}\rangle+ \gamma_5 |\bar{+}\bar{-}\bar{+}\rangle + \varepsilon_5 |\bar{-}\bar{+}\bar{+}\rangle +\theta_5 |\bar{-}\bar{-}\bar{-}\rangle & \;\;\;2 \sqrt{-2+\sqrt{5}} z +\left(-2+\sqrt{5} \right) x & x & 2 \sqrt{-2+\sqrt{5}} z +\left(-2+\sqrt{5} \right) x & x & 2 \sqrt{-2+\sqrt{5}} z +\left(-2+\sqrt{5} \right) x & x\\
6 & 4 \sqrt{2}-1& \left[R(\pi/8) \otimes \openone\right] \frac{1}{\sqrt{2}} \left(|11\rangle_{AC} - |00 \rangle_{AC}\right) & z & x & -\openone & -\openone & z & x\\
7 & 20/3 & \alpha_7 | \tilde{+} \tilde{+} \tilde{+} \rangle + \theta_7 | \tilde{-} \tilde{-} \tilde{-} \rangle & \frac{2\sqrt{2}}{3} z -\frac{1}{3} x & x & \frac{2\sqrt{2}}{3} z -\frac{1}{3} x & x & \frac{2\sqrt{2}}{3} z -\frac{1}{3} x & x\\
8 & 20/3 & \alpha_8 | \tilde{+} \tilde{+} \breve{+} \rangle + \beta_8 | \tilde{+} \tilde{+} \breve{-}\rangle + \eta_8| \tilde{-} \tilde{-} \breve{+} \rangle & \frac{2 \sqrt{2}}{3} z -\frac{1}{3} x & x & \frac{2 \sqrt{2}}{3} z -\frac{1}{3} x & x & z & x\\
9 & 4 \sqrt{2}& \left[R(3 \pi/8) \otimes \openone\right] \frac{1}{\sqrt{2}} \left(|11\rangle_{AB} - |00 \rangle_{AB}\right) & z & x & z & x & -\openone & -\openone\\
10 & 4 & \text{None} & \openone & \openone & \openone & \openone & \openone & \openone\\
11 & 4 \sqrt{2}& \left[R(- \pi/8) \otimes \openone\right] \frac{1}{\sqrt{2}} \left(|00\rangle_{AB} + |11 \rangle_{AB}\right) & z & x & z & x & -\openone & \openone\\
12 & 4 \sqrt{2}& \left[R(\pi/8) \otimes \openone\right] \frac{1}{\sqrt{2}} \left(|00\rangle_{AB} + |11 \rangle_{AB}\right) & z & x & z & x & \openone & -\openone\\
13 & 4 \sqrt{2}& \left[R(\pi/8) \otimes \openone\right] \frac{1}{\sqrt{2}} \left(|11\rangle_{AB} - |00 \rangle_{AB}\right) & z & x & z & x & \openone & -\openone\\
14 & 4 \sqrt{2}& \left[R(\pi/8) \otimes \openone\right] \frac{1}{\sqrt{2}} \left(|11\rangle_{AB} - |00 \rangle_{AB}\right) & z & x & z & x & \openone & -\openone\\
15 & 6 & \left[R(-11\pi/12) \otimes \openone \otimes R(7\pi/12)\right] \frac{1}{\sqrt{2}} \left(|111\rangle - |000 \rangle\right)& \frac{\sqrt{3}}{2} z - \frac{1}{2} x & x & z & x & \frac{\sqrt{3}}{2} z - \frac{1}{2} x & x\\
16 & 6.12883 & \beta_{16} |001\rangle+ \gamma_{16} |010\rangle + \varepsilon_{16} |100\rangle +\theta_{16} |111\rangle & 0.14443 x + 0.98951 z & -0.95407 x - 0.29960 z & -0.61413 x -0.78920 z & -0.61413 x + 0.78920 z & -0.61413 x -0.78920 z& -0.61413 x + 0.78920 z\\
17 & 4 \sqrt{2}& \left[R(\pi/8) \otimes \openone\right] \frac{1}{\sqrt{2}} \left(|00\rangle_{AB} + |11 \rangle_{AB}\right) & z & x & z & x & -\openone & -\openone\\
18 & 2 (7-\sqrt{17}) & \left[R(s) \otimes \openone \otimes \openone \right] \left( \beta_{18} |001\rangle + \gamma_{18} |010\rangle + \epsilon_{18} |100\rangle + \theta_{18} |111 \rangle \right) & g x + \sqrt{1-g^2} z & x & z & x & z & x\\
19 & 5.7829 & \alpha_{19} |000\rangle + \gamma_{19} |010\rangle +\theta_{19} |111\rangle & 0.91209 x + 0.40999 z & -0.91209 x + 0.40999 z & -0.96547 x-0.26053 z & -0.51030 x +0.85999 z & z & x\\
20 & 6 \sqrt{2}-2& \left[R(- 3 \pi/8) \otimes \openone\right] \frac{1}{\sqrt{2}} \left(|00\rangle_{BC} + |11\rangle_{BC}\right) & -\openone & -\openone & z & x & z & x\\
21 & 5.95539 & \alpha_{21} |000\rangle + \beta_{21} |001\rangle + \gamma_{21} |010\rangle + \varepsilon_{21} |100\rangle +\theta_{21}|111\rangle & -0.75982 x-0.6501 z & -0.39634 x+0.91810 z & -0.85421 x+0.51993 z & -0.24405 x-0.96976 z & -0.75016 x+0.66126 z & 0.37339 x+0.92767 z\\
22 & 6.19794 & \alpha_{22} |000\rangle + \beta_{22} |001\rangle + \gamma_{22} |010\rangle + \varepsilon_{22} |100\rangle + \theta_{22} |111\rangle & -0.25333 x + 0.96738 z & 0.99937 x + 0.03540 z & -0.25333 x +0.96738 z & 0.99937 x + 0.03540 z & -0.25333 x + 0.96738 z & 0.99937 x + 0.03540 z\\
23 & \frac{3}{2}\left(\sqrt{17}-1\right) & \alpha_{23} |\hat{+} \hat{+}\rangle_{AB} + \theta_{23} |\hat{-} \hat{-}\rangle_{AB} & f x + \sqrt{1-f^2} z & x & f x + \sqrt{1-f^2} z & x & \openone & \openone\\
24 & 7.94016 & \alpha_{24}|000\rangle + \beta_{24} |001\rangle+ \gamma_{24} |010\rangle + \varepsilon_{24}|100\rangle+\theta_{24} |111\rangle & -0.22334 x+0.97474 z & 0.99876 x-0.04988 z & 0.99439 x+0.10577 z & -0.11740 x+0.99309 z & -0.99439 x+0.10577 z & 0.11740 x+0.99309 z\\
25 & 6.82421 & \alpha_{25}|000\rangle + \beta_{25} |001\rangle + \gamma_{25} |010\rangle + \varepsilon_{25}|100\rangle +\theta_{25} |111\rangle & -0.99611 x-0.08807 z & 0.46076 x+0.88752 z & 0.95922 x-0.28268 z & -0.12828 x+0.99174 z & 0.38106 x+0.92455 z & -0.9949 x+0.10067 z\\
26 & 1+4 \sqrt{3}& \frac{1}{\sqrt{6}} \left( |001 \rangle + |010\rangle - |100\rangle \right) + \frac{1}{\sqrt{2}} |111 \rangle & z & x & z & x & z & x\\
27 & 6.95465 & \alpha_{27} |000\rangle + \beta_{27} |001\rangle + \gamma_{27} |010\rangle + \varepsilon_{27} |100\rangle + \theta_{27} |111\rangle & 0.62948 x+0.77702 z & -0.91661 x+0.39978 z & 0.99744 x+0.07149 z & -0.27833 x+0.96049 z & -0.99744 x+0.0715 z & 0.27833 x+0.96049 z\\
28 & 9.90976 & \alpha_{28} |000\rangle + \beta_{28} |001\rangle + \gamma_{28} |010\rangle + \varepsilon_{28} |100\rangle + \theta_{28} |111\rangle & -0.09603 x+0.99538 z & 0.82667 x+0.56269 z & -0.97402 x+0.22645 z & 0.31271 x+0.94985 z & -0.97402 x+0.22645 z & 0.31271 x+0.94985 z\\
29 & 8 \sqrt{2}-2& \left[R(3 \pi/8) \otimes \openone\right] \frac{1}{\sqrt{2}} \left(|11\rangle_{BC} - |00\rangle_{BC}\right) & -\openone & -\openone & z & x & z & x\\
30 & 8 \sqrt{2}-2& \left[R(- 3 \pi/8) \otimes \openone\right] \frac{1}{\sqrt{2}} \left(|00\rangle_{BC} + |11 \rangle_{BC}\right) & -\openone & -\openone & z & x & z & x\\
31 & 7.80425 & \alpha_{31}|000\rangle + \beta_{31} |001\rangle + \gamma_{31}|010\rangle + \varepsilon_{31} |100\rangle +\theta_{31} |111\rangle & -0.96503 x+0.26212 z & -0.78685 x-0.61714 z & 0.4513 x-0.89237 z & 0.72669 x+0.68697 z & 0.77334 x+0.634 z & -0.97264 x+0.23231 z\\
32 & 8.15161 & \alpha_{32} |000\rangle+ \beta_{32}|001\rangle + \gamma_{32} |010\rangle + \varepsilon_{32} |100\rangle +\theta_{32} |111\rangle & 0.76388 x+0.64535 z & 0.92421 x-0.38189 z & -0.88654 x+0.46265 z & -0.58198 x-0.81321 z & 0.51193 x+0.85903 z & -0.96475 x+0.26319 z\\
33 & 9.78988 & \alpha_{33} |000\rangle + \beta_{33} |001\rangle + \gamma_{33} |010\rangle + \varepsilon_{33} |100\rangle + \theta_{33} |111\rangle & 0.48263 x + 0.87582 z & 0.92087 x - 0.38987 z & 0.48263 x + 0.87583 z & 0.92087 x -0.38987 z & 0.48263 x + 0.87583 z & 0.92087 x -0.38987 z\\
34 & 8.25142 & \beta_{34} |001\rangle + \gamma_{34} |010\rangle + \varepsilon_{34} |100\rangle +\theta_{34} |111\rangle
& -0.84549 x -0.53400 z & 0.845489 x -0.53400 z & -0.96354 x + 0.26755 z & 0.38069 x + 0.92470 z & -0.38069 x + 0.92470 z & 0.96354 x+ 0.26755 z\\
35 & 7.85524 & \beta_{35} |001\rangle + \gamma_{35} |010\rangle + \varepsilon_{35} |100\rangle +\theta_{35} |111\rangle & -0.08798 x -0.99612 z & -0.98249 x + 0.18629 z & -0.98249 x + 0.18629 z & -0.08798 x -0.99612 z & 0.85829 x + 0.51317 z & 0.85829 x -0.51317 z\\
36 & 9.46139 & \alpha_{36} |000\rangle + \beta_{36} |001\rangle + \gamma_{36} |010\rangle + \varepsilon_{36} |100\rangle + \theta_{36} |111\rangle & 0.45381 x-0.8911 z & 0.31791 x+0.94812 z & 0.93502 x+0.35459 z & -0.33145 x+0.94347 z & -0.93502 x+0.35459 z & 0.33145 x+0.94347 z\\
37 & 8 \sqrt{2}-2& \left[R(- \pi/8) \otimes \openone\right] \frac{1}{\sqrt{2}} \left(|11\rangle_{BC} - |00 \rangle_{BC}\right) & -\openone & \openone & z & x & z & x\\
38 & 8 \sqrt{2}-2& \left[R(3 \pi/8) \otimes \openone\right] \frac{1}{\sqrt{2}} \left(|00\rangle_{BC} + |11 \rangle_{BC}\right) & -\openone & \openone & z & x & z & x\\
39 & 9.32530 & \alpha_{39} |000\rangle + \beta_{39} |001\rangle + \gamma_{39} |010\rangle + \varepsilon_{39} |100\rangle + \theta_{39} |111\rangle & -0.04834 x -0.99883 z & -0.99683 x + 0.07953 z & -0.04834 x -0.99883 z & -0.99683 x + 0.07953 z & -0.04834 x -0.99883 z & -0.99683 x + 0.07952 z\\
40 & 8.12983 & \alpha_{40} |000\rangle + \gamma_{40} |010\rangle + \theta_{40}|111\rangle & -0.83322 x + 0.55295 z & 0.83322 x + 0.55295 z & 0.98663 x + 0.16298 z & -0.45177 x + 0.89214 z & z & x\\
41 & 10.3680 & \alpha_{41} |000\rangle + \beta_{41} |001\rangle + \gamma_{41} |010\rangle + \varepsilon_{41} |100\rangle +\theta_{41} |111\rangle & 0.14341 x+0.98966 z & 0.91775 x+0.39717 z & -0.14341 x+0.98966 z & -0.9178 x+0.39717 z & -0.85698 x+0.51535 z & 0.57098 x+0.82097 z\\
42 & 13.0471 & \alpha_{42} |000\rangle + \beta_{42} |001\rangle + \gamma_{42} |010\rangle + \varepsilon_{42} |100\rangle +\theta_{42} |111\rangle & -0.02213 x +0.99976 z & 0.97622 x + 0.21677 z& 0.02213 x + 0.99976 z & -0.97622 x + 0.21677 z& 0.09442 x + 0.99553 z & -0.96162 x + 0.27439 z\\
43 & 8 \sqrt{2}& \left[R(\pi/8) \otimes \openone\right] \frac{1}{\sqrt{2}} \left(|11\rangle_{AB} - |00 \rangle_{AB}\right) & z & x & z & x & -\openone & -\openone\\
44 & 12 \sqrt{2}-4& \left[R(3 \pi/8) \otimes \openone\right] \frac{1}{\sqrt{2}} \left(|00\rangle_{BC} + |11 \rangle_{BC}\right) & -\openone & -\openone & z & x & z & x\\
45 & 12 \sqrt{2}-4& \left[R(-3 \pi/8) \otimes \openone\right] \frac{1}{\sqrt{2}} \left(|11\rangle_{BC} - |00 \rangle_{BC}\right) & -\openone & -\openone & z & x & z & x\\
46 & 12.9852 & \beta_{46} |00\rangle_{AB} +\theta_{46} |11\rangle_{AB} & -0.41094 x-0.91166 z & -0.95751 x+0.28841 z & -0.41094 x-0.91166 z & -0.95751 x+0.28841 z & -\openone & -\openone\\
\hline \hline
\end{tabular}}
\end{center}
\end{table*}
\end{turnpage}
\end{widetext}


\begin{table*}[tb]
\begin{center}
\caption{Coefficients of some states in Table \ref{Table4}.}
\label{Table5}
\begin{tabular}{CCCCCCC}
\hline \hline
\text{\'Sliwa class} & \alpha_i & \beta_i & \gamma_i & \varepsilon_i & \eta_i & \vartheta_i \\
\hline
5  & -\left(-\frac{3}{2}+\frac{7}{2\sqrt{5}} \right)^{1/2} &  \left( \frac{1}{2}-\frac{1}{2\sqrt{5}} \right)^{1/2} & \left(\frac{1}{2}-\frac{1}{2\sqrt{5}} \right)^{1/2} & \left(\frac{1}{2}-\frac{1}{2\sqrt{5}} \right)^{1/2} & & -\left( 1-\frac{2}{\sqrt{5}} \right)^{1/2} \\
7  & \left(\frac{7}{36}+\frac{i \sqrt{2}}{9} \right)^{1/4} & & & & & \left(\frac{7}{36}-\frac{i \sqrt{2}}{9} \right)^{1/4} \\
8  & \frac{\sqrt{2}-4 i}{18} &  \frac{2 i}{3} & & &  \frac{1}{\sqrt{2}} & \\
18 & & -\left[\frac{3}{34} \left(17-3 \sqrt{17}\right) \right]^{1/2} & -\left[\frac{3}{34} \left(17-3 \sqrt{17}\right) \right]^{1/2}& -\frac{1}{2} \left(\frac{31}{\sqrt{17}}-7 \right)^{1/2} & &  \left(\frac{5}{4 \sqrt{17}}-\frac{1}{4} \right)^{1/2} \\
23 & \left[\frac{1}{34} \left(17+\sqrt{17}\right) \right]^{1/2} & & & & & -\left[\frac{1}{34} \left(17-\sqrt{17}\right) \right]^{1/2} \\
16 &            & -0.531856  & -0.531856  & -0.610556  &             & 0.247951  \\
19 & -0.695851  &            & -0.206046  &            &             & 0.687994  \\
21 & 0.0447806  & 0.582265   & 0.565528   & -0.0104097 &             & 0.582265  \\
22 & 0.161337   & 0.421319   & 0.421319   & 0.421319   &             & -0.664411 \\
24 & 0.0475268  & 0.482385   & -0.482385  & 0.343373   &             & 0.643774  \\
25 & -0.392434  & 0.371905   & -0.405923  & 0.399439   &             & 0.619159  \\
27 & 0.701755   & -0.0769086 & 0.076917   & -0.204369  &             & -0.673752 \\
28 & 0.116339   & -0.385379  & -0.385379  & -0.627855  &             & 0.54335   \\
31 & -0.0543542 & 0.421644   & -0.553275  & 0.419715   &             & 0.580507  \\
32 & -0.165302  & 0.411439   & 0.523486   & -0.290699  &             & 0.666971  \\
33 & -0.024223  & 0.452197   & 0.452197   & 0.452197   &             & -0.621262 \\
34 &            & -0.607267  & -0.607267  & 0.195560   &             & -0.473508 \\
35 &            & 0.488723   & -0.613386  & 0.613387   &             & 0.093079  \\
36 & 0.0891467  & 0.632511   & -0.632513  & 0.236068   &             & 0.369029  \\
39 & 0.177347   & 0.522594   & 0.522594   & 0.522594   &             & -0.386311 \\
40 & 0.747562   &            & 0.091742   &            &             & 0.657825  \\
41 & -0.165909  & 0.269314   & -0.474753  & 0.474753   &             & 0.670196  \\
42 & -0.069128  & 0.436957   & -0.437182  & 0.437182   &             & 0.649642  \\
46 &            & 0.756041   &            &            &             & -0.654524 \\
\hline \hline
\end{tabular}
\end{center}
\end{table*}


\begin{table*}[tb]
\begin{center}
\caption{Quantifiers of the tripartite and bipartite entanglement of the states, and incompatibility of the local measurements in Table \ref{Table4}. $N_{ABC}$ is the tripartite negativity (a value 0 indicates no tripartite entanglement, a value 1 indicates GHZ tripartite entanglement), $C_{AB},C_{AC},C_{BC}$ are the qubit-qubit concurrences (0 indicates no bipartite entanglement, 1 indicates maximum bipartite entanglement), and $I(A,a),I(B,b),I(C,c)$ are the quantifiers of incompatibility (0 indicates compatibility, 1 indicates maximum incompatibility).}
\label{Table6}
\begin{tabular}{CCCCCCCCCC}
\hline \hline
\text{\'Sliwa class} & N_{ABC} & C_{AB} & C_{AC} & C_{BC} & \text{Entanglement class} & I(A,a) & I(B,b) & I(C,c) & \text{Incompatibility class} \\
\hline
  1 & 0 & 0 & 0 & 0 & 0 & 0 & 0 & 0 & 0 \\
  2 & 1 & 0 & 0 & 0 & 11 & 1 & 1 & 1 & 11 \\
  3 & 0 & 1 & 0 & 0 & 2 & 1 & 1 & 0 & 4 \\
  4 & 0 & 0 & 0 & 1 & 2 & 0 & 1 & 1 & 4 \\
  5 & 0.933882 & 0.268581 & 0.268581 & 0.268581 & 5 & 0.982759 & 0.982759 & 0.982759 & 7 \\
  6 & 0 & 0 & 1 & 0 & 2 & 1 & 0 & 1 & 4 \\
  7 & 1 & 0 & 0 & 0 & 11 & 0.964724 & 0.964724 & 0.964724 & 7 \\
  8 & 0.980561 & 0.333333 & 0 & 0 & 9 & 0.964724 & 0.964724 & 1 & 9 \\
  9 & 0 & 1 & 0 & 0 & 2 & 1 & 1 & 0 & 4 \\
  10 & 0 & 0 & 0 & 0 & 0 & 0 & 0 & 0 & 0 \\
  11 & 0 & 1 & 0 & 0 & 2 & 1 & 1 & 0 & 4 \\
  12 & 0 & 1 & 0 & 0 & 2 & 1 & 1 & 0 & 4 \\
  13 & 0 & 1 & 0 & 0 & 2 & 1 & 1 & 0 & 4 \\
  14 & 0 & 1 & 0 & 0 & 2 & 1 & 1 & 0 & 4 \\
  15 & 1 & 0 & 0 & 0 & 11 & 0.914836 & 1 & 0.914836 & 9 \\
  16 & 0.963789 & 0.385708 & 0.385708 & 0.262966 & 4 & 0.937797 & 0.981286 & 0.981286 & 6 \\
  17 & 0 & 1 & 0 & 0 & 2 & 1 & 1 & 0 & 4 \\
  18 & 0.915977 & 0.165537 & 0.165537 & 0.651125 & 4 & 0.713778 & 1 & 1 & 10 \\
  19 & 0.984687 & 0 & 0.283517 & 0 & 9 & 0.831756 & 0.977506 & 1 & 8 \\
  20 & 0 & 0 & 0 & 1 & 2 & 0 & 1 & 1 & 4 \\
  21 & 0.942194 & 0.668329 & 0.648551 & 0.648551 & 4 & 0.972541 & 0.972541 & 0.964724 & 6 \\
  22 & 0.961728 & 0.296515 & 0.296515 & 0.296515 & 5 & 0.985224 & 0.985224 & 0.985224 & 7 \\
  23 & 0 & 0.970142 & 0 & 0 & 1 & 0.985183 & 0.985183 & 0 & 2 \\
  24 & 0.968662 & 0.296208 & 0.296208 & 0.0654725 & 4 & 0.976974 & 0.999958 & 0.999958 & 6 \\
  25 & 0.948037 & 0.504699 & 0.527644 & 0.522772 & 3 & 0.899609 & 0.947022 & 0.974378 & 5 \\
  26 & 0.942809 & 0.244017 & 0.244017 & 0.244017 & 5 & 1 & 1 & 1 & 11 \\
  27 & 0.979794 & 0.135081 & 0.135081 & 0.287211 & 4 & 0.977897 & 0.986566 & 0.986567 & 6 \\
  28 & 0.963674 & 0.142217 & 0.142217 & 0.405471 & 4 & 0.922076 & 0.997574 & 0.997574 & 6 \\
  29 & 0 & 0 & 0 & 1 & 2 & 0 & 1 & 1 & 4 \\
  30 & 0 & 0 & 0 & 1 & 2 & 0 & 1 & 1 & 4 \\
  31 & 0.984264 & 0.0679139 & 0.295241 & 0.066422 & 3 & 0.870692 & 0.974559 & 0.866792 & 5 \\
  32 & 0.94546 & 0.329232 & 0.509299 & 0.224655 & 3 & 0.929528 & 0.994058 & 0.97765 & 5 \\
  33 & 0.983258 & 0.155835 & 0.155835 & 0.155835 & 5 & 0.996785 & 0.996785 & 0.996785 & 7 \\
  34 & 0.947084 & 0.337577 & 0.337577 & 0.552346 & 4 & 0.939217 & 0.99567 & 0.99567 & 6 \\
  35 & 0.935 & 0.661507 & 0.485366 & 0.485364 & 4 & 0.997022 & 0.997022 & 0.924734 & 6 \\
  36 & 0.917495 & 0.180609 & 0.180612 & 0.629359 & 3 & 0.806014 & 0.999817 & 0.999817 & 6 \\
  37 & 0 & 0 & 0 & 1 & 2 & 0 & 1 & 1 & 4 \\
  38 & 0 & 0 & 0 & 1 & 2 & 0 & 1 & 1 & 4 \\
  39 & 0.970318 & 0.197647 & 0.197647 & 0.197647 & 5 & 0.999705 & 0.999705 & 0.999705 & 7 \\
  40 & 0.988444 & 0 & 0.1207 & 0 & 9 & 0.951145 & 0.971647 & 1 & 8 \\
  41 & 0.946929 & 0.239827 & 0.440842 & 0.440842 & 4 & 0.904916 & 0.904916 & 0.998673 & 6 \\
  42 & 0.972206 & 0.206077 & 0.206518 & 0.20651 & 4 & 0.988316 & 0.988316 & 0.989814 & 6 \\
  43 & 0 & 1 & 0 & 0 & 2 & 1 & 1 & 0 & 4 \\
  44 & 0 & 0 & 0 & 1 & 2 & 0 & 1 & 1 & 4 \\
  45 & 0 & 0 & 0 & 1 & 2 & 0 & 1 & 1 & 4 \\
  46 & 0 & 0.989694 & 0 & 0 & 1 & 0.994818 & 0.994818 & 0 & 2 \\
 \hline \hline
\end{tabular}
\end{center}
\end{table*}


\subsection{Classification of maximum nonlocality}


Table \ref{Table7} contains the classification of the maximum nonlocality in the $(3,2,2)$ scenario according to the simplest class of entanglement and of incompatibility required to reach it. Simplest means with less entanglement and less incompatibility. We observe that there are 15 classes of nonlocality.
Some cases, for example, the nonlocality class $(2,4)$ corresponding to the entanglement class 2 and the incompatibility class 4, maximally violate many inequalities (16 inequalities). In contrast, there are classes of nonlocality like, e.g., $(5,11)$, which maximally violate only one inequality. A detailed analysis of Table \ref{Table7} and the conclusions that can be extracted from it is presented in the next section.


\begin{table*}[t]
\begin{center}
\caption{Classification of the maximum nonlocality of the $(3,2,2)$ scenario according to the simplest combination of entanglement and incompatibility required. The intersection between each class of entanglement and each class of incompatibility is occupied by the tight Bell inequality maximally violated by that combination. Bell inequalities are numbered as in Table \ref{Table1}.}
\label{Table7}
\begin{tabular}{CCCCCCCCCCCCC}
\hline \hline
\multirow{2}{3.0cm}{\centering \text{Incompatibility class}} & \multicolumn{12}{c}{\centering \text{Entanglement class}}\\
\cline{2-13} & \multicolumn{1}{c}{0} & \multicolumn{1}{c}{1} & \multicolumn{1}{c}{2} & \multicolumn{1}{c}{3} & \multicolumn{1}{c}{4} & \multicolumn{1}{c}{5} & \multicolumn{1}{c}{6} & \multicolumn{1}{c}{7} & \multicolumn{1}{c}{8} & \multicolumn{1}{c}{9} & \multicolumn{1}{c}{10} & \multicolumn{1}{c}{11} \\
\hline
0  & 1,10 & & & & & & & & & & & \\
1  & & & & & & & & & & & & \\
2  & & 23,46 & & & & & & & & & & \\
3  & & & & & & & & & & & & \\
\multirow{3}{*}{4} & & & 3,4,6,9,11,12 & & & & & & & & & \\
                   & & & 13,14,17,20,29,30 & & & & & & & & & \\
                   & & & 37,38,43,44,45 & & & & & & & & & \\
5  & & & & 25,31,32 & & & & & & & & \\
\multirow{2}{*}{6}& & & & 36 & 16,21,24,27 & & & & & & & \\
                  & & & &    & 34,35,41 & & & & & & & \\
7  & & & & & 42 & 5,22,33,39 & & & & & & 7\\
8  & & & & & & & & & & 19,40 & & \\
9  & & & & & & & & & & 8 & & 15 \\
10 & & & & & 18 & & & & & & & \\
11 & & & & & & 26 & & & & & & 2 \\
\hline \hline
\end{tabular}
\end{center}
\end{table*}


\section{Analysis and interesting cases}


\subsection{General observations}


The following observations can be made in the light of Table \ref{Table7}:

(i) When we focus on the simplest combinations of entanglement and incompatibility, we observe that only some combinations produce maximum nonlocality. This is vividly illustrated in Table \ref{Table7}, where we see that almost all cases occur in the diagonal of the Table. This shows that there should be a tuned balance between entanglement and incompatibility in order to produce maximum nonlocality. This balance tuning deserves further investigation. The few cases which are located out of the diagonal correspond to either maximum entanglement (inequalities 7 and 15) or maximum incompatibility (inequalities 18 and 26).

(ii) Bipartite nonmaximally entangled states maximally violate two tight Bell inequalities in the $(3,2,2)$ scenario (inequalities 23 and 46). And, moreover, in both cases using nonmaximum incompatibility. This contrasts with the $(2,2,2)$ scenario, where nonmaximally entangled states cannot maximally violate the only tight Bell inequality \cite{SW87,Tsirelson93}. Inequality 23 will receive special attention in the next subsection.

(iii) Bipartite maximally entangled states and maximum incompatibility (i.e., the same resources needed for maximally violating the CHSH inequality) are enough for maximally violating a high number of tight Bell inequalities in the the $(3,2,2)$ scenario (inequalities 3, 4, 6, 9, 11--14, 17, 20, 29, 30, 37, 38, and 43--45). This shows that maximum tripartite nonlocality does not require either tripartite entanglement or tripartite incompatibility. That maximum tripartite nonlocality does not require tripartite quantum resources can also be seen from the biseparable upper bounds computed independently by Vallins, Sainz, and Liang \cite{VSL16}.

(iv) There are no cases in between (ii) and (iii). This means that for maximum nonlocality maximum bipartite entanglement always goes with maximum incompatibility, and nonmaximum bipartite entanglement always goes with nonmaximum incompatibility. This is the simplest example of the tuned balance mentioned in (i).

(v) There is no tight Bell inequality maximally violated by the W state. This shows that maximum (in a sense) tripartite nonlocality and maximum tripartite entanglement are different concepts. However, there are many inequalities maximally violated by states of the classes 3, 4, and 5, i.e., by W-like states in which, respectively, none, two, or three of the concurrences are equal. In 11 of these cases, $N_{ABC}$ is larger than the one of the W state, in three cases (inequalities 5, 18, and 21) $N_{ABC}$ is smaller than the one of the W state, and in one case (inequality 26) $N_{ABC}$ is exactly equal to the one of the W state (but the state has smaller qubit-qubit concurrences than the W state). Inequality 26 will receive special attention in a later subsection.

(vi) Only a small number of inequalities are only maximally violated by GHZ states (inequalities 2, 7, and 15). In fact, only the maximum violation of the well-known Mermin inequality \cite{Mermin90} (\'Sliwa inequality 2) requires maximum incompatibility for all parties.

(vii) Besides the Mermin inequality (inequality 2), which is only maximally violated by the GHZ state, there is only one inequality whose maximum violation requires maximum incompatibility for the three parties: inequality 26. As we have said, it will receive special attention in a later subsection.


\subsection{\'Sliwa inequality 23}


The set of almost quantum correlations \cite{NGHA15} is defined as those correlations for which there exists a quantum state such that for all permutations of operators for $n$-parties leave the statistics invariant. In the $(2,2,2)$ scenario, this definition is equivalent to the set of correlations arising when we assume that all correlations allowed by level $Q_{1+AB}$ of the NPA hierarchy \cite{NPA07,NPA08} are physical. This is a supra quantum set, but it is attractive for, at least, two reasons: It satisfies a long list of reasonable principles and, unlike the corresponding quantum set, the almost quantum set for a given Bell inequality scenario is easy to characterize. Similarly, while deciding whether or not some correlations belong to the corresponding quantum set is conjectured to be undecidable \cite{NGHA15}, this is not the case for the almost quantum set. If quantum theory is correct, then there is no way to falsify almost quantum correlations in experiments. Therefore, either one constructs an explicit theory which gives rise to the almost quantum set and identifies an experiment that falsifies quantum theory or points out why almost quantum correlations are nonphysical.

Inequality 23 sheds some light on this problem. On one hand, as it has also been pointed out by Vallins, Sainz, and Liang \cite{VSL16}, for two out of the 46 inequalities, inequalities 23 and 41, the set of almost quantum correlations allows for larger than quantum values ($4.7754$ and $10.3735$, respectively), which means that almost quantum correlations predict larger than quantum nonlocality in these cases. The interesting thing is that, as shown in Table \ref{Table4}, the only way the quantum maximum of inequality 23 can be achieved with qubits is when the third observer measures the identity. In other words, when he does not measure anything and always outputs $1$ no matter which the setting is. This is the way quantum theory achieves $\frac{3}{2}(\sqrt{17}-1) \approx 4.6847$, while almost quantum correlations predict that $4.7754$ can be reached.

This would suggest that almost quantum correlations should also give higher than quantum violation for the following bipartite Bell inequality, which is obtained from inequality 23 by making $C=c=\openone$:
\begin{equation}
2 A + 2 B + a b - a B - A b - 3 A B \le 4.
\label{ineq}
\end{equation}
However, for inequality (\ref{ineq}) the almost quantum maximum agrees with the quantum one [which is, obviously, $\frac{3}{2}(\sqrt{17}-1)$]. This points out that either the almost quantum set is nonphysical because it violates a fundamental principle, or that theories producing almost quantum correlations should allow a new kind of tripartite entanglement which does not exist in quantum theory. To illustrate how this second option would work, it is useful to consider the Mermin inequality \cite{Mermin90} in the case where the third observer measures the identity. Then, the Mermin inequality reduces to the CHSH inequality \cite{CHSH69}. Both the CHSH and the Mermin inequalities share the same local bound, but the quantum violation of the Mermin inequality is larger than the one of the CHSH inequality. The reason for that is that there is genuinely tripartite quantum entanglement. In contrast, our results show that the gap between quantum and almost quantum correlations in inequality 23 cannot be explained through genuinely tripartite quantum entanglement, but requires a different kind of genuinely tripartite nonquantum entanglement.

However, we conjecture that the answer to the puzzle is not this option but the first one. Moreover, we conjecture that in both the bi- and tripartite cases, the quantum value $\frac{3}{2}(\sqrt{17}-1)$ saturates the exclusivity principle \cite{Cabello13,Yan13,AFLS15,Cabello15} and that, therefore, the reason why the almost quantum set is nonphysical is because it violates the exclusivity principle. Specifically, we conjecture that for inequality 23, quantum theory saturates the exclusivity principle when we consider a suitable extension of the $(3,2,2)$ scenario. That is, a subsequent set of experiments in addition to those strictly needed for testing inequality 23. An example of such an extension for the $(n,2,2)$ scenario can be found in Ref.\ \cite{Cabello15}.


\subsection{\'Sliwa inequality 26}


The work of Greenberger, Horne, and Zeilinger \cite{GHZ89} led to identify the Mermin inequality \cite{Mermin90} (inequality 2) and opened the door to the exploration of multipartite entanglement and nonlocality, both theoretically and experimentally, both for basic science and applications. Although there are 46 tight Bell inequalities in the $(3,2,2)$ scenario considered by GHZ, all the experimental exploration of tripartite nonlocality has been essentially focused on the Mermin inequality \cite{PBDWZ00,EMFLHYPBPSRGLWJR14,HSHMMVMNRJ14,LZJHDBBR14}. We may wonder whether there is any other tight Bell inequality whose quantum violation is worth being experimentally investigated. There are different properties which point out some inequalities as interesting. For example, inequality 7 is interesting because it belongs \cite{LRBPBG15} to a family of $n$-partite Bell inequalities which extends the CHSH inequality \cite{WW01,ZB02}, inequality 10 is interesting because it is violated by some nonsignaling theories but not by quantum theory \cite{ABBAGP10}, and inequality 15 is interesting for its efficiency for detecting nonlocality \cite{PMS16}. However, none of these reasons justify new experiments.

On the other hand, in this paper we are interested in maximum nonlocalities which are special from the point of view of the combination of resources needed to achieve them. Maximum nonlocalities requiring Hardy, Bell, or GHZ states can be taken as examples of nonlocalities that are special in this sense. However, there is no experimental challenge in testing them, since this has already been done.

Interestingly, if one assumes that incompatibility is the most basic resource of the two, and that the role of entanglement is extracting the best from incompatibility, then a natural question is whether there is any tight Bell inequality whose maximum quantum violation requires maximum incompatibility for the three parties but not a GHZ state. The answer is yes. There is only one case, inequality 26. Unlike Mermin's case, in this case there is bipartite entanglement among all pairs of qubits. Still, the maximum quantum violation is almost as high as in the case of the Mermin inequality. Moreover, it is the only tight Bell inequality maximally violated by a state that has the same tripartite negativity as the W state (but it is not the W state) and whose preparation constitutes an experimental challenge by itself. This definitely makes inequality 26 worthy of experimental tests. Furthermore, it identifies it as a potential source of new applications.

Not only that. Inequality 26 also singles out a tripartite entangled state which is interesting by itself. On one hand, the classification in Table \ref{Table7} reveals that W-like states, like the one which maximally violates inequality 26, are the most frequent tripartite entangled states leading to maximum nonlocality. On the other hand, when studying entanglement, certain states with nice properties emerge. For instance, the state that maximizes the 4-qubit hyperdeterminant \cite{AMM10} is
$|\psi_4\rangle =\frac{1}{\sqrt{6}} \left( |0001\rangle + |0010\rangle + |0100\rangle + |1000\rangle \right) + \frac{1}{\sqrt{3}} |1111\rangle$.
Other interesting states are also of the form $|\Psi \rangle = p \left( |00 \cdots 01\rangle + |00 \cdots 10\rangle + \cdots + |01 \cdots 00\rangle + |10 \cdots 00\rangle \right) + q|11\cdots 11\rangle$ \cite{AMM10}.
For these states, a typical question is whether there is a tight Bell inequality that is only maximally violated by them. Inequality 26 provides one, as it is only maximally violated by the state
\begin{equation}
\label{state}
|\psi\rangle =\frac{1}{\sqrt{6}} \left(|001\rangle+|010\rangle-|100\rangle\right)+\frac{1}{\sqrt{2}}|111\rangle,
\end{equation}
which is essentially of the form $|\Psi \rangle$. This finding leads us to the conjecture that, the same way there is a family of tight Bell inequalities in the $(n,2,2)$ scenario which is only maximally violated by GHZ states \cite{Mermin90,Ardehali92,BK93}, there should be a family of tight Bell inequalities generalizing inequality 26 and only maximally violated by states generalizing state (\ref{state}). An interesting challenge is to identify this family. On the other hand, it is very likely that many families of tight Bell inequalities in the $(n,2,2)$ scenario are only maximally violated by generalized W-like states.


\section{Conclusions}


We have classified the ways quantum theory combines entanglement and nonlocality in order to produce maximum nonlocality for the 46 classes of facets of the local polytope in the $(3,2,2)$ scenario. This classification is important for two reasons. One of them is for understanding how entanglement and incompatibility should combine to produce maximum nonlocality and for identifying combinations that escaped previous research but are potential sources of insight, experimental challenges, novel applications, and interesting theoretical questions. A priori, it could have happened that nothing new came out from this analysis. However, the list of conclusions reached and observations pointed out in the previous section made it worth the effort.

Nevertheless, the main reason why our work is interesting is because it provides the starting point for solving the main open problem in the field, namely, identifying the principle that singles out quantum correlations. In this respect, we have shown that inequality 23 may provide the key to solve the problem. The principle that explains the maximum quantum violation of inequality 23 will very likely be the principle that governs quantum correlations. Of course, that principle should also explain the quantum maxima of all 46 inequalities. We hope that this work stimulates further research in this direction.


\begin{acknowledgments}
We thank Daniel Cavalcanti, Dardo Goyeneche, Teiko Heinosaari, Matty Hoban, Jos\'e Ignacio Latorre, Yeong-Cherng Liang, Ravishankar Ramanathan, Ana Bel\'en Sainz, Debasis Sarkar, and Karol \.Zyczkowski for useful conversations.
We are also grateful to Matthias Kleinmann and Tam\'as V\'ertesi for checking some of the calculations, to Antonio J. L\'opez-Tarrida for his comments on the manuscript, and to Yeong-Cherng Liang for pointing out an error in a previous version.
This work was supported by Project No.\ FIS2014-60843-P, ``Advanced Quantum Information'' (MINECO, Spain), with FEDER funds,
by the FQXi Large Grant ``The Observer Observed: A Bayesian Route to the Reconstruction of Quantum Theory,''
and by the project ``Photonic Quantum Information'' (Knut and Alice Wallenberg Foundation, Sweden).
Z.-P.\ X.\ is supported by the Natural Science Foundation of China (Grant No.\ 11475089).
\end{acknowledgments}




\begin{thebibliography}{99}


\bibitem{Bell64}
 J. S. Bell,
 On the Einstein Podolsky Rosen paradox,
 Physics (Long Island City, NY) \textbf{1}, 195 (1964).


\bibitem{Gisin91}
 N. Gisin,
 Bell's inequality holds for all non-product states,
 \href{http://dx.doi.org/10.1016/0375-9601(91)90805-I}{Phys. Lett. A \textbf{154}, 201 (1991).}

\bibitem{PR92}
 S. Popescu and D. Rohrlich,
 Which states violate Bell's inequality maximally?,
 \href{http://dx.doi.org/10.1016/0375-9601(92)90819-8}{Phys. Lett. A \textbf{169}, 411 (1992).}

\bibitem{Werner89}
 R. F. Werner,
 Quantum states with Einstein-Podolsky-Rosen correlations admitting a hidden-variable model,
 \href{http://dx.doi.org/10.1103/PhysRevA.40.4277}{Phys. Rev. A \textbf{40}, 4277 (1989).}

\bibitem{AGT06}
 A. Ac\'{\i}n, N. Gisin, and B. Toner,
 Grothendieck’s constant and local models for noisy entangled quantum states,
 \href{https://doi.org/10.1103/PhysRevA.73.062105}{Phys. Rev. A \textbf{73}, 062105 (2006).}

\bibitem{Vertesi08}
 T. V\'ertesi,
 More efficient Bell inequalities for Werner states,
 \href{https://doi.org/10.1103/PhysRevA.78.032112}{Phys. Rev. A \textbf{78}, 032112 (2008).}

\bibitem{BFFHB16}
 J. Bowles, J. Francfort, M. Fillettaz, F. Hirsch, and N. Brunner,
 Genuinely Multipartite Entangled Quantum States with Fully Local Hidden Variable Models and Hidden Multipartite Nonlocality,
 \href{http://dx.doi.org/10.1103/PhysRevLett.107.210403}{Phys. Rev. Lett. \textbf{116}, 130401 (2016).}

\bibitem{CGRS16}
 D. Cavalcanti, L. Guerini, R. Rabelo, and P. Skrzypczyk,
 General Method for Constructing Local Hidden Variable Models for Entangled Quantum States,
 \href{https://doi.org/10.1103/PhysRevLett.117.190401}{Phys. Rev. Lett. \textbf{117}, 190401 (2016).}

\bibitem{HQVPB16}
 F. Hirsch, M. T. Quintino, T. V\'ertesi, M. F. Pusey, and N. Brunner,
 Algorithmic Construction of Local Hidden Variable Models for Entangled Quantum States,
 \href{https://doi.org/10.1103/PhysRevLett.117.190402}{Phys. Rev. Lett. \textbf{117}, 190402 (2016).}


\bibitem{KC85}
 L. A. Khalfin and B. S. Tsirelson,
 Quantum and quasi-classical analogs of Bell inequalities,
 \href{http://www.tau.ac.il/~tsirel/download/khts85.pdf}{in
 {\em Symposium on the Foundations of Modern Physics: 50 Years of the Einstein-Podolsky-Rosen Experiment},
 edited by P. J. Lahti and P. Mittelstaedt
 (World Scientific, Singapore, 1985), p.~441.}

\bibitem{WPF09}
 M. M. Wolf, D. P\'erez-Garc\'{\i}a, and C. Fern\'andez,
 Measurements Incompatible in Quantum Theory Cannot Be Measured Jointly in Any Other No-Signaling Theory,
 \href{https://doi.org/10.1103/PhysRevLett.103.230402}{Phys. Rev. Lett. \textbf{103}, 230402 (2009).}

\bibitem{QBHB16}
 M. T. Quintino, J. Bowles, F. Hirsch, and N. Brunner,
 Incompatible quantum measurements admitting a local-hidden-variable model,
 \href{https://doi.org/10.1103/PhysRevA.93.052115}{Phys. Rev. A \textbf{93}, 052115 (2016).}


\bibitem{CHSH69}
 J. F. Clauser, M. A. Horne, A. Shimony, and R. A. Holt,
 Proposed Experiment to Test Local Hidden-Variable Theories,
\href{http://dx.doi.org/10.1103/PhysRevLett.23.880}{Phys. Rev. Lett. \textbf{23}, 880 (1969).}

\bibitem{SW87}
 S. J. Summers and R. F. Werner,
 Maximal violation of Bell's inequalities is generic in quantum field theory,
 \href{http://dx.doi.org/10.1007/BF01207366}{Commun. Math. Phys. \textbf{110}, 247 (1987).} See Theorem 2.3.

\bibitem{Tsirelson93}
 B. S. Tsirelson,
 Some results and problems on quantum Bell-type inequalities,
 \href{http://www.tau.ac.il/~tsirel/download/hadron.pdf}{Hadronic J. Suppl. \textbf{8}, 329 (1993).}


\bibitem{BL95}
 P. Busch and P. J. Lahti,
 The complementarity of quantum observables: Theory and experiments,
 \href{https://doi.org/10.1007/BF02743814}{Rivista del Nuovo Cimento \textbf{18}, 1 (1995).}

\bibitem{HMZ16}
 T. Heinosaari, T. Miyadera, and M. Ziman,
 An invitation to quantum incompatibility,
 \href{https://doi.org/10.1088/1751-8113/49/12/123001}{J. Phys. A: Math. Theor. \textbf{49}, 123001 (2016).}


\bibitem{Froissart81}
 M. Froissart,
 Constructive generalization of Bell's inequalities,
 \href{https://doi.org/10.1007/BF02903286}{Nuovo Cimento B \textbf{64}, 241 (1981).}

\bibitem{CG04}
 D. Collins and N. Gisin,
 A relevant two qubit Bell inequality inequivalent to the CHSH inequality,
 \href{http://dx.doi.org/10.1088/0305-4470/37/5/021}{J. Phys. A: Math. Gen. \textbf{37}, 1775 (2004).}

\bibitem{PV10}
 K. F. P\'al and T. V\'ertesi,
 Maximal violation of a bipartite three-setting, two-outcome Bell inequality using infinite-dimensional quantum systems,
 \href{http://dx.doi.org/10.1103/PhysRevA.82.022116}{Phys. Rev. A \textbf{82}, 022116 (2010).}

\bibitem{VW11}
 T. Vidick and S. Wehner,
 More nonlocality with less entanglement,
 \href{http://dx.doi.org/10.1103/PhysRevA.83.052310}{Phys. Rev. A \textbf{83}, 052310 (2011).}


\bibitem{CGLMP02}
 D. Collins, N. Gisin, N. Linden, S. Massar, and S. Popescu,
 Bell Inequalities for Arbitrarily High-Dimensional Systems,
 \href{http://dx.doi.org/10.1103/PhysRevLett.88.040404}{Phys. Rev. Lett. \textbf{88}, 040404 (2002).}

\bibitem{ADGL02}
 A. Ac\'{\i}n, T. Durt, N. Gisin, and J. I. Latorre,
 Quantum nonlocality in two three-level systems,
 \href{http://dx.doi.org/10.1103/PhysRevA.65.052325}{Phys. Rev. A \textbf{65}, 052325 (2002).}

\bibitem{ZG08}
 S. Zohren and R. D. Gill,
 Maximal Violation of the Collins-Gisin-Linden-Massar-Popescu Inequality for Infinite Dimensional States,
 \href{http://dx.doi.org/10.1103/PhysRevLett.100.120406}{Phys. Rev. Lett. \textbf{100}, 120406 (2008).}

\bibitem{SLK04}
 W. Son, J. Lee, and M. S. Kim,
 $d$-outcome measurement for nonlocality test,
 \href{http://dx.doi.org/10.1088/0305-4470/37/49/009}{J. Phys. A: Math. Gen. \textbf{37}, 11897 (2004).}


\bibitem{GHZ89}
 D. M. Greenberger, M. A. Horne, and A. Zeilinger,
 Going beyond Bell's theorem,
 in
 {\em Bell's Theorem, Quantum Theory, and Cconceptions of the Universe},
 edited by M. Kafatos
 (Kluwer, Dordrecht, Holland, 1989), p.~69.

\bibitem{Mermin90}
 N. D. Mermin,
 Extreme quantum entanglement in a superposition of macroscopically distinct states,
 \href{https://doi.org/10.1103/PhysRevLett.65.1838}{Phys. Rev. Lett. \textbf{65}, 1838 (1990).}

\bibitem{Ardehali92}
 M. Ardehali,
 Bell inequalities with a magnitude of violation that grows exponentially with the number of particles,
 \href{https://doi.org/10.1103/PhysRevA.46.5375}{Phys. Rev. A \textbf{46}, 5375 (1992).}

\bibitem{BK93}
 A. V. Belinsky and D. N. Klyshko,
 \begin{otherlanguage*}{russian}
 Интерференция света и теорема Белла,
 \end{otherlanguage*}
 \href{https://doi.org/10.3367/UFNr.0163.199308a.0001}{Usp. Fiz. Nauk \textbf{63}, 1 (1993).}
 [English version:
 Interference of light and Bell's theorem,
 \href{https://doi.org/10.1070/PU1993v036n08ABEH002299}{Phys. Usp. \textbf{36}, 653 (1993).}]

\bibitem{GC08}
 O. G{\"u}hne and A. Cabello,
 Generalized Ardehali-Bell inequalities for graph states,
 \href{https://doi.org/10.1103/PhysRevA.77.032108}{Phys. Rev. A \textbf{77}, 032108 (2008).}

\bibitem{CGR08}
 A. Cabello, O. G{\"u}hne, and D. Rodr\'{\i}guez,
 Mermin inequalities for perfect correlations,
 \href{https://doi.org/10.1103/PhysRevA.77.062106}{Phys. Rev. A \textbf{77}, 062106 (2008).}


\bibitem{PS01}
 I. Pitowsky and K. Svozil,
 Optimal tests of quantum nonlocality,
 \href{https://doi.org/10.1103/PhysRevA.64.014102}{Phys. Rev. A \textbf{64}, 014102 (2001).}

\bibitem{Sliwa03}
 C. \'Sliwa,
 Symmetries of the Bell correlation inequalities,
 \href{https://arxiv.org/abs/quant-ph/0305190}{\eprint{arXiv:quant-ph/0305190}.} The published version
 [\href{http://dx.doi.org/10.1016/S0375-9601(03)01115-0}{Phys. Lett. A \textbf{317}, 165 (2003)}] does not contain the table with the 46 Bell inequalities.

\bibitem{Masanes03}
 L. Masanes,
 Extremal quantum correlations for $N$ parties with two dichotomic observables per site,
 \href{https://arxiv.org/abs/quant-ph/0512100}{\eprint{arXiv:quant-ph/0512100}.}


\bibitem{PPKSWZ09}
 M. Paw{\l}owski, T. Paterek, D. Kaszlikowski, V. Scarani, A. Winter, and M. \.{Z}ukowski,
 Information causality as a physical principle,
 \href{http://dx.doi.org/10.1038/nature08400}{Nature (London) \textbf{461}, 1101 (2009).}

\bibitem{GWAN11}
 R. Gallego, L. E. W\"urflinger, A. Ac\'{\i}n, and M. Navascu\'es,
 Quantum Correlations Require Multipartite Information Principles,
 \href{http://dx.doi.org/10.1103/PhysRevLett.107.210403}{Phys. Rev. Lett. \textbf{107}, 210403 (2011).}

\bibitem{YCATS12}
 T. H. Yang, D. Cavalcanti, M. L. Almeida, C. Teo, and V. Scarani,
 Information-causality and extremal tripartite correlations,
 \href{http://dx.doi.org/10.1088/1367-2630/14/1/013061}{New J. Phys. \textbf{14}, 013061 (2012).}

\bibitem{vanDam99}
 W. van Dam
 {\em Nonlocality \& Communication Complexity},
 Ph.D. thesis, University of Oxford, 1999.

\bibitem{vanDam13}
 W. van Dam
 Implausible consequences of superstrong nonlocality,
 \href{http://dx.doi.org/10.1007/s11047-012-9353-6}{Nat. Comput. \textbf{12}, 9 (2013).}

\bibitem{NW09}
 M. Navascu\'es and H. Wunderlich,
 A glance beyond the quantum model,
 \href{http://dx.doi.org/10.1098/rspa.2009.0453}{Proc. Royal Soc. A \textbf{466}, 881 (2009).}

\bibitem{FSABCLA13}
 T. Fritz, A. B. Sainz, R. Augusiak, J. Bohr Brask, R. Chaves, A. Leverrier, and A. Ac\'{\i}n,
 Local orthogonality: A multipartite principle for correlations,
 \href{http://dx.doi.org/10.1038/ncomms3263}{Nat. Commun. \textbf{4}, 2263 (2013).}

\bibitem{VSL16}
 J. Vallins, A. B. Sainz, and Y.-C. Liang,
 Almost quantum correlations and their refinements in tripartite Bell scenarios,
 \href{http://arxiv.org/abs/1608.05641}{\eprint{arXiv:1608.05641 [quant-ph]}.}

\bibitem{NGHA15}
 M. Navascu\'es, Y. Guryanova, M. J. Hoban, and A. Ac\'{\i}n,
 Almost quantum correlations,
 \href{http://dx.doi.org/10.1038/ncomms7288}{Nat. Commun. \textbf{6}, 6288 (2015).}


\bibitem{NPA07}
 M. Navascu\'es, S. Pironio, and A. Ac\'{\i}n,
 Bounding the Set of Quantum Correlations,
 \href{https://doi.org/10.1103/PhysRevLett.98.010401}{Phys. Rev. Lett. \textbf{98}, 010401 (2007).}

\bibitem{NPA08}
 M. Navascu\'es, S. Pironio, and A. Ac\'{\i}n,
 A convergent hierarchy of semidefinite programs characterizing the set of quantum correlations,
 \href{https://10.1088/1367-2630/10/7/073013}{New J. Phys. \textbf{10}, 073013 (2008).}


\bibitem{SG08}
 C. Sab\'{\i}n and G. Garc\'{\i}a-Alcaine,
 A classification of entanglement in three-qubit systems,
 \href{https://doi.org/10.1140/epjd/e2008-00112-5}{Eur. Phys. J. D \textbf{48}, 435 (2008).}

 \bibitem{AACJLT00}
 A. Ac\'{\i}n, A. Andrianov, L. Costa, E. Jan\'e, J. I. Latorre, and R. Tarrach,
 Generalized Schmidt Decomposition and Classification of Three-Quantum-Bit States,
 \href{https://doi.org/10.1103/PhysRevLett.85.1560}{Phys. Rev. Lett. \textbf{85}, 1560 (2000).}

\bibitem{AAJT01}
 A. Ac\'{\i}n, A. Andrianov, E. Jan\'e, and R. Tarrach,
 Three-qubit pure-state canonical forms,
 \href{https://doi.org/10.1088/0305-4470/34/35/301}{J. Phys. A: Math. Gen. \textbf{34}, 6725 (2001).}

\bibitem{Vidal00}
 G. Vidal,
 Entanglement monotones,
 \href{https://doi.org/10.1080/09500340008244048}{J. Mod. Opt. \textbf{47}, 355 (2000).}

\bibitem{HW97}
 S. Hill and W. K. Wootters,
 Entanglement of a Pair of Quantum Bits,
 \href{https://doi.org/10.1103/PhysRevLett.78.5022}{Phys. Rev. Lett. \textbf{78}, 5022 (1997).}

\bibitem{Hardy93}
 L. Hardy,
 Nonlocality for two particles without inequalities for almost all entangled states,
 \href{https://doi.org/10.1103/PhysRevLett.71.1665}{Phys. Rev. Lett. \textbf{71}, 1665 (1993).}

\bibitem{BMR92}
 S. L. Braunstein, A. Mann, and M. Revzen,
 Maximal violation of Bell inequalities for mixed states,
 \href{https://doi.org/10.1103/PhysRevLett.68.3259}{Phys. Rev. Lett. \textbf{68}, 3259 (1992).}

\bibitem{DVC00}
 S. D\"urr, G. Vidal, and J. I. Cirac,
 Three qubits can be entangled in two inequivalent ways,
 \href{https://doi.org/10.1103/PhysRevA.62.062314}{Phys. Rev. A \textbf{62}, 062314 (2000).}

\bibitem{PB03}
 M. Plesch and V. Bu\v{z}ek,
 Entangled graphs: Bipartite entanglement in multiqubit systems,
 \href{https://doi.org/10.1103/PhysRevA.67.012322}{Phys. Rev. A \textbf{67}, 012322 (2003).}


\bibitem{HKR15}
 T. Heinosaari, J. Kiukas, and D. Reitzner,
 Noise robustness of the incompatibility of quantum measurements,
 \href{https://doi.org/10.1103/PhysRevA.92.022115}{Phys. Rev. A \textbf{92}, 022115 (2015).}


\bibitem{Cabello13}
 A. Cabello,
 Simple Explanation of the Quantum Violation of a Fundamental Inequality,
 \href{http://dx.doi.org/10.1103/PhysRevLett.110.060402}{Phys. Rev. Lett. \textbf{110}, 060402 (2013).}

\bibitem{Yan13}
 B. Yan,
 Quantum Correlations are Tightly Bound by the Exclusivity Principle,
 \href{https://doi.org/10.1103/PhysRevLett.110.260406}{Phys. Rev. Lett. \textbf{110}, 260406 (2013).}

\bibitem{AFLS15}
 A. Ac\'{\i}n, T. Fritz, A. Leverrier, and A. B. Sainz,
 A Combinatorial Approach to Nonlocality and Contextuality,
 \href{http://dx.doi.org/10.1103/PhysRevLett.112.040401}{Comm. Math. Phys. \textbf{334}, 533 (2015).}

\bibitem{Cabello15}
 A. Cabello,
 Simple Explanation of the Quantum Limits of Genuine $n$-Body Nonlocality,
 \href{http://dx.doi.org/10.1103/PhysRevLett.114.220402}{Phys. Rev. Lett. \textbf{114}, 220402 (2015).}


\bibitem{PBDWZ00}
 J.-W. Pan, D. Bouwmeester, M. Daniell1, H. Weinfurter, and A. Zeilinger,
 Experimental test of quantum nonlocality in three-photon Greenberger–Horne–Zeilinger entanglement,
 \href{https://doi.org/10.1038/35000514}{Nature (London) \textbf{403}, 515 (2000).}

\bibitem{EMFLHYPBPSRGLWJR14}
 C. Erven,
 E. Meyer-Scott,
 K. Fisher,
 J. Lavoie,
 B. L. Higgins,
 Z. Yan,
 C. J. Pugh,
 J.-P. Bourgoin,
 R. Prevedel,
 L. K. Shalm,
 L. Richards,
 N. Gigov,
 R. Laflamme,
 G. Weihs,
 T. Jennewein, and
 K. J. Resch,
 Experimental three-photon quantum nonlocality under strict locality conditions,
 \href{http://dx.doi.org/10.1038/nphoton.2014.50.html}{Nat. Photon. \textbf{8}, 292 (2014).}

\bibitem{HSHMMVMNRJ14}
 D. R. Hamel,
 L. K. Shalm,
 H. H\"ubel,
 A. J. Miller,
 F. Marsili,
 V. B. Verma,
 R. P. Mirin,
 S. W. Nam,
 K. J. Resch, and
 T. Jennewein,
 Direct generation of three-photon polarization entanglement,
 \href{http://dx.doi.org/10.1038/nphoton.2014.218}{Nat. Photon. \textbf{8}, 801 (2014).}

\bibitem{LZJHDBBR14}
 B. P. Lanyon, M. Zwerger, P. Jurcevic, C. Hempel, W. D\"ur, H. J. Briegel, R. Blatt, and C. F. Roos,
 Experimental Violation of Multipartite Bell Inequalities with Trapped Ions,
 \href{https://doi.org/10.1103/PhysRevLett.112.100403}{Phys. Rev. Lett. \textbf{112}, 100403 (2014).}


\bibitem{LRBPBG15}
 Y.-C. Liang, D. Rosset, J.-D. Bancal, G. P\"utz, T. J. Barnea, and N. Gisin,
 Family of Bell-like Inequalities as Device-Independent Witnesses for Entanglement Depth,
 \href{https://doi.org/10.1103/PhysRevLett.114.190401}{Phys. Rev. Lett. \textbf{114}, 190401 (2015).}

\bibitem{WW01}
 R. F. Werner and M. M. Wolf,
 All-multipartite Bell-correlation inequalities for two dichotomic observables per site,
 \href{https://doi.org/10.1103/PhysRevA.64.032112}{Phys. Rev. A \textbf{64}, 032112 (2001).}

\bibitem{ZB02}
 M. \.{Z}ukowski and \v{C}. Brukner,
 Bell's Theorem for General $N$-Qubit States,
 \href{https://doi.org/10.1103/PhysRevLett.88.210401}{Phys. Rev. Lett. \textbf{88}, 210401 (2002).}

\bibitem{ABBAGP10}
 M. L. Almeida, J.-D. Bancal, N. Brunner, A. Ac\'{\i}n, N. Gisin, and S. Pironio,
 Guess Your Neighbor's Input: A Multipartite Nonlocal Game with No Quantum Advantage,
 \href{http://dx.doi.org/10.1103/PhysRevLett.104.230404}{Phys. Rev. Lett. \textbf{104}, 230404 (2010).}

\bibitem{PMS16}
 B. Paul, K. Mukherjee, and D. Sarkar,
 Nonlocality of three-qubit Greenberger-Horne-Zeilinger-symmetric states,
 \href{https://doi.org/10.1103/PhysRevA.94.032101}{Phys. Rev. A \textbf{94}, 032101 (2016).}

\bibitem{AMM10}
 M. Aulbach, D. Markham, and M. Murao,
 The maximally entangled symmetric state in terms of the geometric measure,
 \href{https://doi.org/10.1088/1367-2630/12/7/073025}{New J. Phys. \textbf{12}, 073025 (2010).}


\end{thebibliography}
\end{document}